\documentclass[reprint,amsmath,amssymb,superscriptaddress]{revtex4-1}
\usepackage[T1]{fontenc}
\usepackage[utf8]{inputenc}
\usepackage{graphicx,amsmath}
\usepackage{xcolor}

\newcommand{\FigLandscape}{Supplementary Figure~S2}
\newcommand{\FigProba}{Supplementary Figure~S3}
\newcommand{\FigCap}{Supplementary Figure~S4}

\usepackage{pdfpages}
\makeatletter
\AtBeginDocument{\let\LS@rot\@undefined}
\makeatother
\usepackage{pgffor}

\begin{document}
\title{Formation of porous crystals via viscoelastic phase separation} 

\author{Hideyo Tsurusawa} 
\affiliation{Institute of Industrial Science, University of Tokyo, 4-6-1 Komaba, Meguro-ku, Tokyo 153-8505, Japan}
\author{John Russo}
\affiliation{Institute of Industrial Science, University of Tokyo, 4-6-1 Komaba, Meguro-ku, Tokyo 153-8505, Japan}
\affiliation{ {School of Mathematics, University of Bristol, Bristol BS8 1TW, United Kingdom} }
\author{Mathieu Leocmach}
\affiliation{Univ Lyon, Université Claude Bernard Lyon 1, CNRS, Institut Lumière Matière, F-69622, VILLEURBANNE, France}
\author{Hajime Tanaka  \footnote{e-mail: tanaka@iis.u-tokyo.ac.jp}}
\affiliation{Institute of Industrial Science, University of Tokyo, 4-6-1 Komaba, Meguro-ku, Tokyo 153-8505, Japan}

\begin{abstract}
{\bf
Viscoelastic phase separation of colloidal suspensions can be interrupted to form gels either by glass transition or by crystallization. 
With a new confocal microscopy protocol, we follow the entire kinetics of phase separation, from homogeneous phase to different arrested states.
For the first time in experiments, our results unveil a novel crystallization pathway to sponge-like porous crystal structures. In the early stages, we show that nucleation requires a structural reorganization of the liquid phase, called stress-driven ageing.
Once nucleation starts, we observe that crystallization follows three different routes: 
direct crystallization of the liquid phase, Bergeron process, and Ostwald ripening. 
Nucleation starts inside the reorganised network, but crystals grow past it by direct condensation of the gas phase on their surface, driving liquid evaporation, and producing a network structure different from the original phase separation pattern.
We argue that similar crystal-gel states can be formed in monoatomic and molecular systems if the liquid phase is slow enough to induce viscoelastic phase separation, but fast enough to prevent immediate vitrification. 
This provides a novel pathway to form nano-porous crystals of metals and semiconductors without dealloying, which may be important for catalytic, optical, sensing, and filtration applications. 
}
\end{abstract}
\maketitle


Crystallization plays a fundamental role in many processes occurring in nature, such as ice formation in atmospheric clouds~\cite{glickman2000glossary,morrison2012resilience}, and in technological applications that are at the core of the chemical, pharmaceutical, and food industries.
Many of the properties of crystals, like the shape, spatial arrangement, polymorph type, and size distribution of the crystallites, depend on the conditions at which the nucleation process took place. 
Controlling the early stages of crystallization is thus of fundamental importance in order to obtain in a reproducible manner crystals with the desired properties. 

Classical Nucleation Theory describes the formation of an ordered crystalline nucleus directly from the supersaturated solution. 
But crystallization can be preceded by the formation of dense liquid  droplets as an intermediate step~\cite{ten1997enhancement,SearR,savage2009experimental,vekilov2010two,palberg2014crystallization}.
Understanding the process of crystal formation in mixed-phase systems (composed of gas, liquid, and solid phases) is thus of great importance for a variety of systems, from protein solutions to clouds. 

Colloidal suspensions offer a system where the crystallization process in a mixed-phase environment can be observed with single-particle
resolution, and at the same timescales over which nucleation takes place. In colloids with short-range attractions the gas-liquid transition 
becomes metastable with respect to crystallization~\cite{anderson2002insights,lekkerkerker2011colloids}, and can form gels   
~\cite{poon2002,zaccarelli2007,piazza1994phase,verhaegh1997transient,lu2008gelation}. 
This gel formation process can be regarded as viscoelastic phase separation~\cite{tanaka1999colloid} into a dense liquid phase with slow dynamics and a dilute gas phase with fast dynamics.  
This difference in the viscoelastic properties between the two phases allows the formation of a space-spanning network structure of the liquid phase, even if it is the minority phase. 
The network is fractal and its dynamic structure factor shows a stretched-exponential decay to a finite plateau~\cite{krall1998internal,solomon2001dynamic,romer2000sol}. 
During phase separation, the density of the liquid phase increases towards the glass-transition density, leading to slow glassy dynamics.
At the glass transition point, the percolated network structure is dynamically stabilised by vitrification of the dense liquid phase~\cite{pusey1993dynamics,piazza1994phase,ilett1995phase,verhaegh1997transient,
tanaka1999colloid,foffi2002,buzzaccaro2007sticky,zaccarelli2007,lu2008gelation,
testard2011}.  
More precisely, the dynamical stabilization is due to percolation of locally favoured structures, which are locally stable non-crystalline structures~\cite{royall2008g}.

Unlike the above standard scenario of colloidal gelation, it was shown that phase separation can also be accompanied by crystallization~\cite{poon1999cpm}, if the process takes place below the melting point of one of the two separating phases~\cite{tanaka1985new}, and the phenomena were discussed on the basis of its unique free-energy landscape~\cite{poon1999cpm}. 
Later the possibility of a different class of gels, which are stabilized by crystallization, was suggested by numerical simulations~\cite{soga1999metastable,charbonneau2007systematic,fortini2008crystallization,
perez2011pathways} and observed in experiments~\cite{sabin2012,zhang2012non}.
In Brownian dynamics simulations, Ref.~\cite{fortini2008crystallization} found that crystallization  is enhanced  by  the  metastable  gas-liquid  binodal by  means  of  a  two-stage  crystallization  process. 
Despite being less likely than for purely attractive systems, the possibility of interrupting spinodal decomposition with crystallization was observed also in simulations of charged attractive colloids~\cite{charbonneau2007systematic}, and in both experiments and simulations of oppositely charged colloids~\cite{sanz2008gel,sanz2008out}.

However, there has so far been no experimental single-particle level studies on the \emph{dynamical process} of crystal-gel formation, which makes the microscopic mechanism of crystal-gel formation elusive. 
For confocal microscopy experiments on colloidal gels, colloid-polymer mixtures have so far been used as a model system~\cite{poon2002,lu2008gelation,royall2008g,zhang2012non}. 
The main difficulty consists in the preparation protocol of the unstable suspension into an initial homogeneous state. 
Shaking or applying shear introduces turbulent flows at the beginning of the process, prevents the observation of the initial stages of gelation, and may even alter the selection of the final non-equilibrium arrested state.  
In our experiments we instead succeed in building a novel protocol that allows the time evolution of the system to be observed in a quiescent situation without introducing fluid flows. 
We thus address by single-particle-level observation how the momentum conservation, which is a consequence of the presence of a liquid component, does affect the selection of the kinetic pathway of gelation.

In this Article, we show that the system can undergo a novel crystallization pathway that leads to the formation of interconnected crystalline droplets, different from both vitrified or crystallised liquid network pathways. 
We show that after a spinodal decomposition, crystal nucleation in the liquid network is made possible by stress-driven ageing which releases the stresses that build-up during the network formation due to hydrodynamic interactions, allowing for a local increase in density necessary to accommodate critical nuclei.
Stress-driven ageing is possible only when interparticle attraction is weak enough, otherwise we observe the classical gelation process, in which phase separation is interrupted by vitrification~\cite{verhaegh1997transient,tanaka1999colloid,lu2008gelation}.
We then show that the growth of the crystal seeds occurs by three different routes:
(i) Growth within the dense branches of the gel;
(ii) Bergeron process, analogous to ice formation in mixed phase clouds~\cite{glickman2000glossary,morrison2012resilience}, where ice droplets grow at the expense of the supercooled liquid droplets due to their lower saturated vapour pressure;
(iii) Ostwald ripening.
Differently from the standard crystal-gel scenario, where crystals only nucleate and grow inside the dense branches of the gel, becoming dynamically arrested, we show that the kinetic pathways involving the gas phase (predominantly the Bergeron process, but also Ostwald ripening) play an important role.
We reveal that the final crystal-gel network structure has smoother interface than the ordinary gels formed by vitrification because of this novel crystal growth mechanism.

\begin{figure}[h]
 \includegraphics{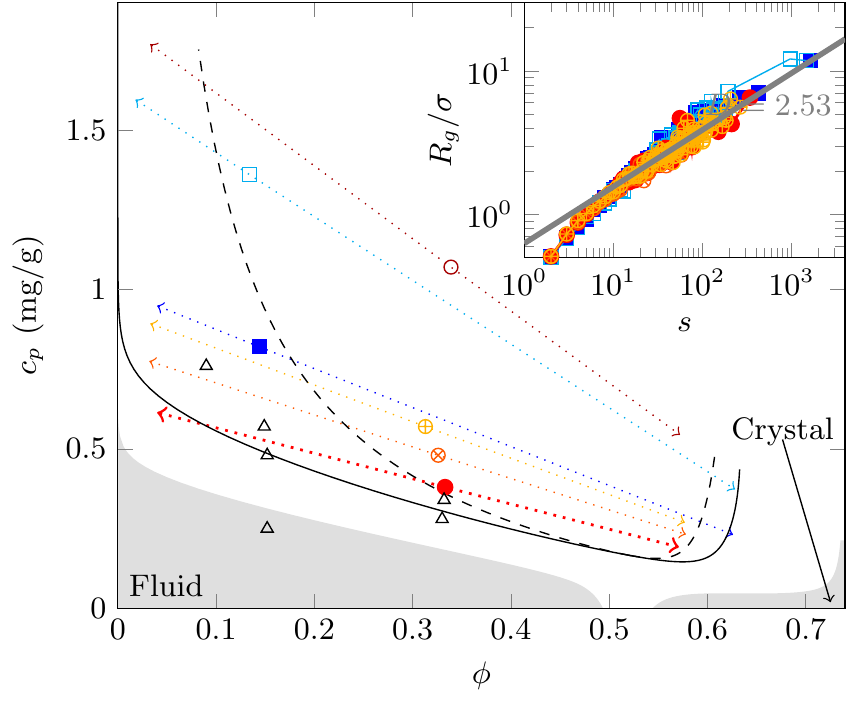}
 \caption{{\bf Phase behaviour.} 
Empty black triangles are experimental points showing fluid behaviour. The other symbols are the experimental points showing gel behaviour, consistently used in all figures. The dotted lines are the gas-liquid tie lines determined experimentally, with arrow heads indicating measured gas and liquid compositions (see Supplementary Information on the procedure to extract the compositions from confocal images).  Grey areas are theoretical equilibrium fluid and crystal regions. The solid and dashed lines are the metastable gas-liquid binodal and spinodal respectively.
\textbf{(Inset)} Radius of gyration ($R_g$) as a function of cluster size ($s$) for all state points at percolation time. The grey line represents the fractal dimension of the random gelation universality class ($D=2.53$).
 }
 \label{fig:phasediagram}
\end{figure}

In our experimental setup, the sample cell is put in contact with a reservoir of a salt solution through a semi-permeable membrane (see Methods).
At $t=0$, the increase in the concentration of salt ions within a few Brownian times ($\tau_{\rm B}$) screens the Coulomb repulsion between colloidal particles, which are then subject to attractive depletion forces, making the system thermodynamically unstable and leading to spinodal decomposition without any harmful flow or drift, as confirmed by Supplementary Movie~1.
In Fig.~\ref{fig:phasediagram}, we superimpose the state points with a theoretical phase diagram (see Supplementary Methods).
The state points which did not phase separate are indicated with open triangles, while the other symbols indicate samples undergoing phase separation on which we focus in the following. 

A typical early stage phase ordering process observed at $\phi\approx 0.33$ and $c_p=0.38$ mg/g can be seen in the beginning of Supplementary Movie~1. 
Due to strong dynamical asymmetry between colloids and the solvent~\cite{tanaka1999colloid}, the colloidal particles start aggregating, 
eventually forming a percolating network. Thus, a dense network (liquid) coexists with freely diffusing monomers (gas).

At percolation, all samples display a similar network topology. 
This is shown for example in the inset of Fig.~\ref{fig:phasediagram}, where we plot the radius of gyration of colloidal clusters ($R_g$) as a function of the cluster size ($s$) at the percolation time, see Methods.
All samples show the same scaling law, $R_{\rm g}\sim s^{1/D}$, which is compatible with the random gelation universality class exponent ($D=2.53$) as shown by the grey line in the inset of Fig.~\ref{fig:phasediagram}. 
Here we do not imply that the formation of the network follows the random gelation universality class, but just that our fractal (or effective) dimension is compatible with it.

\begin{figure}
 \includegraphics[width=\columnwidth]{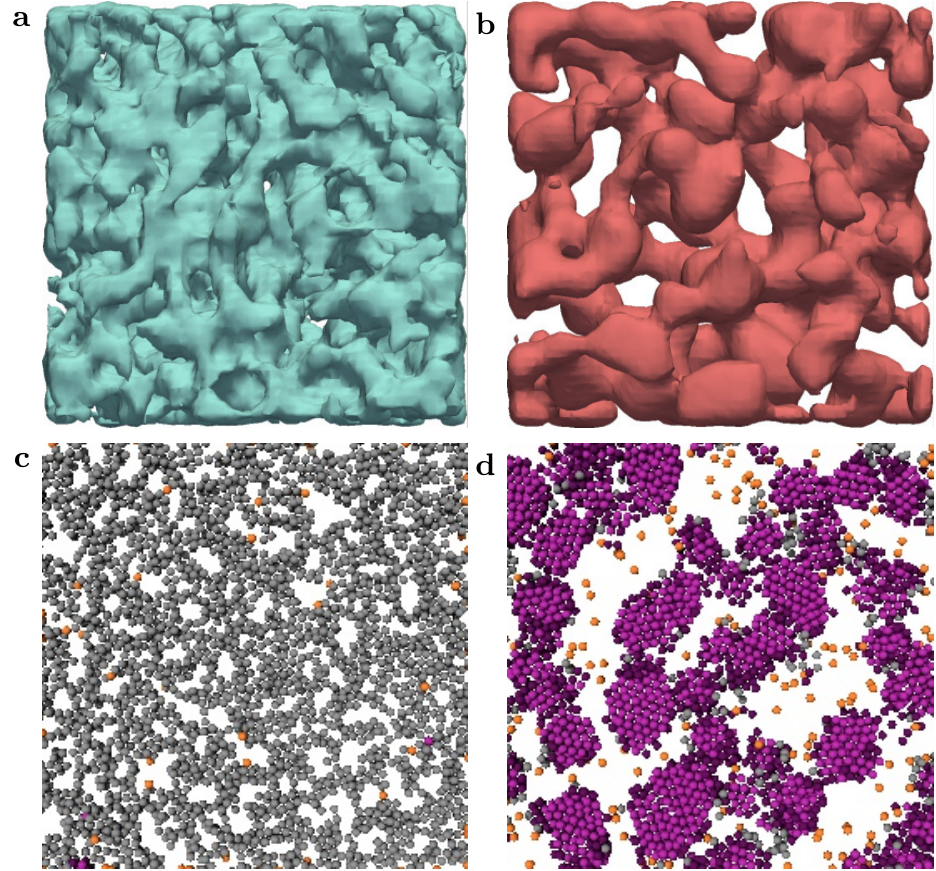}
\caption{{\bf Percolated network structures.} 
{\bf a, b:} the network is represented in a coarse-grained fashion (see Methods).
{\bf c, d:} slabs of 4.5 $\mu$m thickness and 41 $\mu$m width. Particles are drawn to scale and coloured according to their phase. Purple: crystal; dark purple: crystal surface; grey: liquid; orange: gas. 
{\bf a, c:} a gel obtained from the state point $\phi\approx 0.33$ and $c_p=1.07$ mg/g at $t=400$ min after sample preparation. 
{\bf b, d:} a crystal-gel at the same $\phi$ but at lower polymer concentration, $c_p=0.38$ mg/g at $t=468$ min after sample preparation. 
On the top row we can clearly see not only the percolated nature of both networks, but also the difference in the smoothness of the network surface between amorphous and crystal gels. 
We note that very few liquid particles remain in \textbf{d} and the network is almost crystalline. 
} 
\label{fig:network}
\end{figure}

The samples share the same early stages of the phase separation process, but show significantly different behaviours in the later stages.
An example of this is given in Fig.~\ref{fig:network} where we compare the final network structures for two samples at the same colloid volume fraction $\phi\approx 0.33$ but at different polymer concentrations, $c_p=1.07$ mg/g and $c_p=0.38$ mg/g.
In Fig.~\ref{fig:network}a and b, Gaussian filtering is used to depict the network structure as a continuous field (see Methods). 
The comparison shows that high polymer concentration (panel a) produces a network of thin strands whereas low polymer concentration produces a beaded network structure with bigger pores, smoother surfaces, and thicker strands (panel b). 
Fig.~\ref{fig:network}c and d shows the individual particle positions for slabs with a thickness of 5 particle diameters. 
From the figures it is immediately clear that the network strands at low polymer concentration are crystalline (Fig.~\ref{fig:network}d) but not the ones at higher $c_p$ (Fig.~\ref{fig:network}c).

\begin{figure*}
 \includegraphics{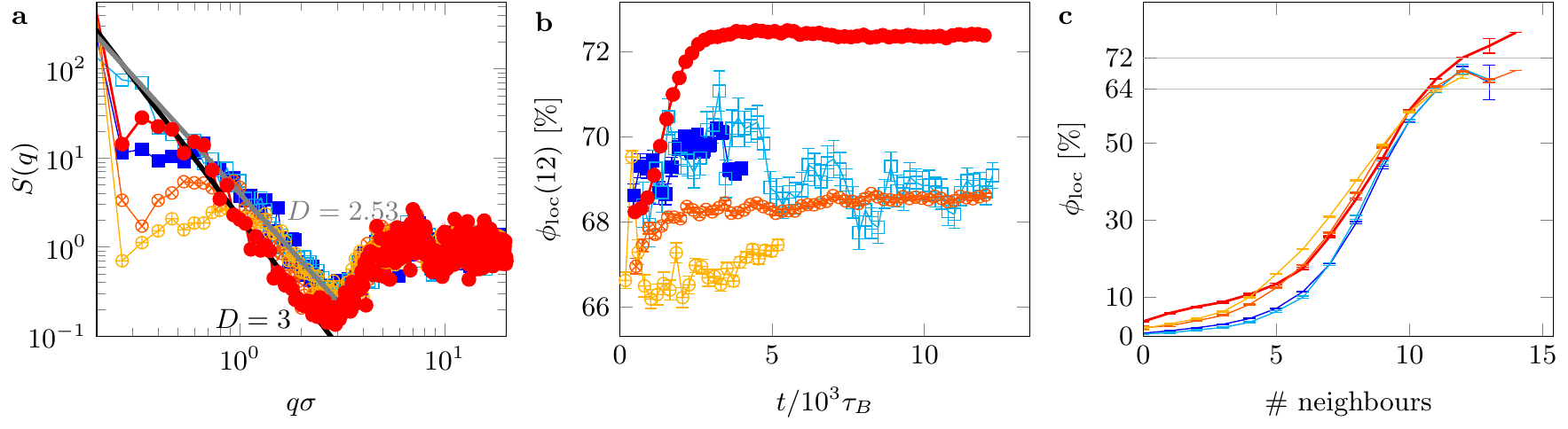}
 \caption{{\bf Structural evolution at late times.} 
{\bf a,} Structure factors for the late stages of the gelation process. The grey and black lines represent respectively random gelation and compact fractal dimensions.
{\bf b,} Late stage of the evolution of the local volume fraction $\phi_{\rm loc}$(12) around colloidal particles having 12 nearest neighbours.
{\bf c,} Local volume fraction $\phi_{\rm loc}$ around a colloidal particle, as a function of the number of nearest neighbours of the particle for the state points in the very late stage. 
The two horizontal straight lines indicate the characteristic volume fractions in attractive systems, crystal ($\sim 0.72$) and glass ($\sim 0.64$).
Samples are shown with the same colour code as in Fig.~\ref{fig:phasediagram}. In particular red circles correspond to the crystallizing sample at $\phi=0.33$ and $c_p=0.38\,$mg/g.
 }
 \label{fig:late_structure}
\end{figure*}

To confirm the generality of these two distinct final structures, we plot in Fig.~\ref{fig:late_structure}a the structure factor for the final stages of the gelation process for all state points.
At low wavenumber $q$, the structure factor displays fractal scaling compatible with the Guinier law, $S(q)\sim q^{-D}$.
But a difference in the fractal (or effective) dimension $D$ between the $\phi\approx 0.33$ and $c_p=0.38$ mg/g state point and other state points is visible.
In fact, while states with high polymer concentration retain the exponent $D=2.5(3)$, which is the random gel universality class exponent, as we also observed in the early stages of the gelation process (see the inset of Fig.~\ref{fig:phasediagram}), the state point with low polymer concentration ($\phi\approx 0.33$ and $c_p=0.38$ mg/g) displays a slope compatible with the exponent $D=3$, which corresponds to compact structures. This is consistent with the smoother surfaces of the beaded network (see Fig.~\ref{fig:network}b). 

In Fig.~\ref{fig:late_structure}b we plot the late stage (after all samples already underwent gas-liquid phase separation) time evolution of the local volume fraction $\phi_\mathrm{loc}(12)$ of closed packed particles (having 12 neighbours, see Methods). Only the state point at $\phi\approx 0.33$ and $c_p = 0.38\,$mg/g shows an increase, up to an asymptotic value that we identify with the composition of the stable crystal phase $\approx 0.72$. We plot the asymptotic average volume fraction as a function of the number of neighbours in Fig.~\ref{fig:late_structure}c. Approaching close packing (12 neighbours), we clearly see two families of curves. The state point at $\phi\approx 0.33$ and $c_p=0.38\,$mg/g reaches close packing at the volume fraction $\approx 0.72$. 
By contrast, all other state points reach close packing at a markedly lower volume fraction, which is indeed close to the volume fraction of the attractive glass state~\cite{pham2002multiple}.
To sum up, at the state point $\phi\approx 0.33$ and $c_p=0.38\,$mg/g the network crystallises whereas other samples stay glassy.
In particular, for state points with $\phi\approx 0.33$, increasing polymer concentration drastically reduces the amount of crystals. 
This is in agreement with the idea of enhanced crystallization rates near metastable critical points~\cite{ten1997enhancement,olmsted1998spinodal} and the results of Refs.~\cite{soga1999metastable,fortini2008crystallization,perez2011pathways},
which speculated two different arrest mechanisms: crystallization at low polymer concentration, and dynamic arrest at high polymer concentrations. 
This is in contradiction with free energy landscape approach~\cite{poon1999cpm}, predicting that all our samples should crystallise after the initial spinodal decomposition (see Supplementary Methods and \FigLandscape). 
This indicates that the selection of the kinetic pathway in a non-equilibrium process cannot be inferred from free energy alone, but is strongly affected by kinetic factors, e.g. the difference in the kinetics between spinodal decomposition and crystallization \cite{tanaka1985new}, and hydrodynamic and mechanical effects (see below).

\begin{figure}
 \includegraphics{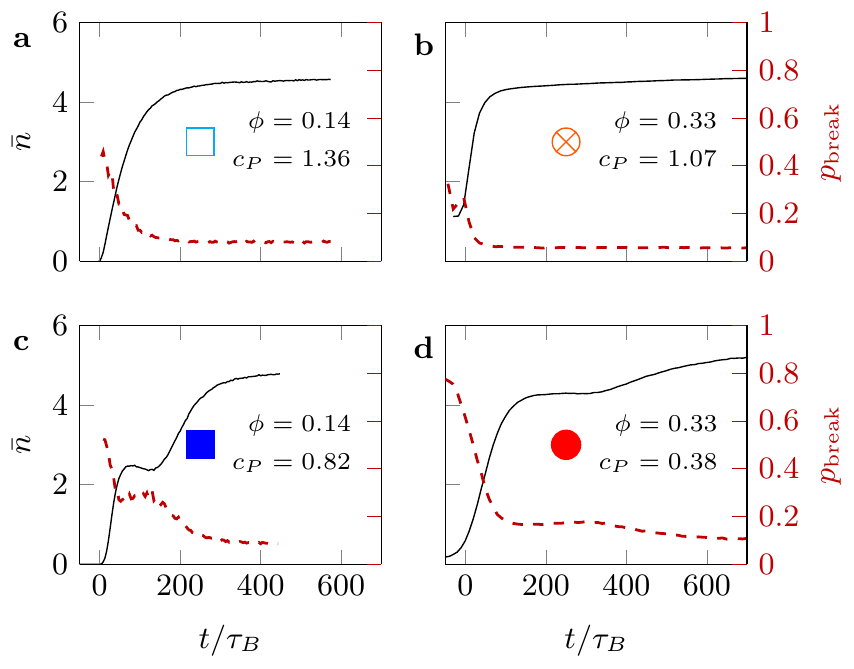}
 \caption{{\bf Stress driven ageing.} Time evolution of the average coordination of colloidal particles $\bar{n}$ (black continuous line) and the bond-breaking probability $p_\text{break}$ (red dashed line) for four gel samples. Symbols relate each sample to its position on the phase diagram in Fig.~\ref{fig:phasediagram}. $p_\text{break}$ is measured as the probability of bond breaking in a time interval of $\Delta t=10$ s. For high polymer concentration, panels {\bf a} and {\bf b}, $p_\text{break}$ decreases monotonically until it reaches a stationary state at long times, with $\bar{n}$ saturating at an average of less than $5$ neighbours. For low polymer concentration, panels {\bf c} and {\bf d}, $p_\text{break}$ follows a similar decay at intermediate times, with $\bar{n}$ less than $5$ neighbours, but which is then followed by a second decay to new configuration, where $\bar{n}$ becomes greater than $5$ neighbours.}
 \label{fig:bond_breaking}
\end{figure}

We argue that the main difference between our samples is the occurrence or absence of stress-driven ageing of the network. First, let us recall that the network formed by the initial spinodal decomposition has thin strands, just a few particles thick (see Fig.~\ref{fig:network}a and c). 
Such morphology can be explained only by taking into account the fundamental role played by hydrodynamic interactions from initial to intermediate times, before percolation is complete~\cite{tanaka2000,tanaka2007spontaneous,furukawa2010key}.
Without hydrodynamic interactions, as often assumed in simulations, particles have the tendency to aggregate in compact structures and subsequently form thick network structures. 
With hydrodynamic interactions, particles first form a transient gel even at very low volume fractions, and the number of nearest neighbours increases only later to minimize the energy of the structure. 
Thus, hydrodynamic interactions lead to the formation of gels that are very far from equilibrium and under a strong thermodynamic driving force towards more stable compact structures. 
The resulting transition from open to more compact networks occurs through the breaking of the bonds that have accumulated more stress. 
This process has been called stress-driven ageing~\cite{tanaka2007spontaneous}, and is characteristic of viscoelastic phase separation~\cite{tanaka1999colloid}. 
Whether the network undergoes such restructuring or not depends on the strength of the bonds, i.e. the polymer concentration.
In Fig.~\ref{fig:bond_breaking} we plot the bond breaking probability, $p_\text{bond}$, and the average coordination, $\bar{n}$, as a function of the time elapsed since salt injection ($t=0$).
At high polymer concentrations (panels \textbf{a}, \textbf{b}), $p_\text{break}$ decreases monotonically, until a stationary state is reached where $\bar{n}<5$.
At low polymer concentrations (panels \textbf{c}, \textbf{d}), the initial decay of $p_\text{break}$ to a network with $\bar{n}<5$ is followed by a secondary decay to a more compact network with $\bar{n}>6$. This means that mechanical stress built up in the first transient network can relax to a more compact network only when bonds are weak enough, i.e. at low polymer concentrations.

It is worth noting that the two-step behaviour of $p_\text{break}$ and $\bar{n}$ is not due to crystallization, as this only starts after the network reorganization. For example, the fraction of crystalline particles for all times displayed in Fig.~\ref{fig:bond_breaking}d is always below $0.3\%$.
The increase in number of bonds after network reorganization is a necessary condition for crystallization, as more compact environments become available to accommodate the critical nucleus size, while in its absence the network forms low-density arrested states (gels).

\begin{figure}
 \includegraphics{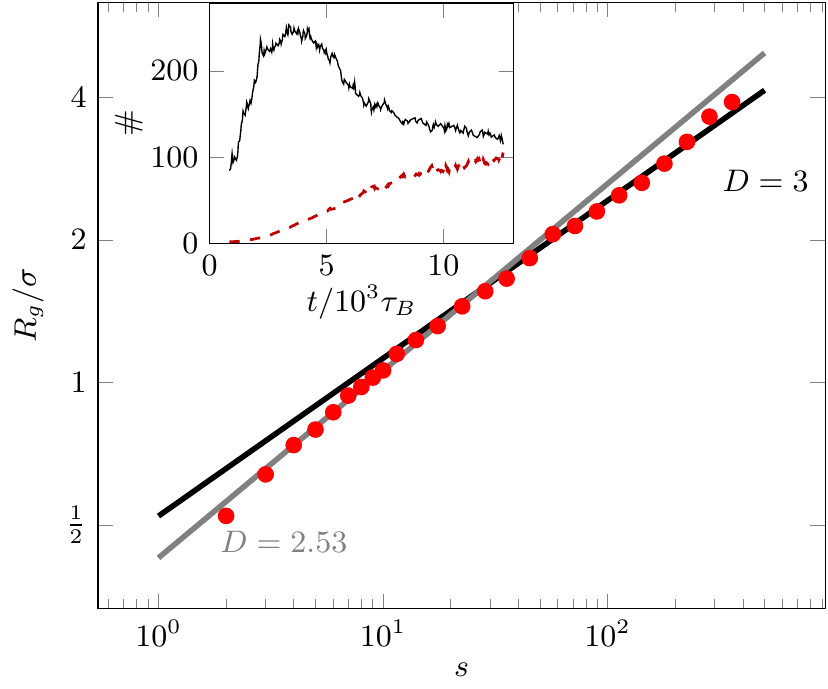}
\caption{{\bf Characterization of the crystal-gel.} 
Radius of gyration of crystalline nuclei at the late stages of gelation for the state point $\phi\approx 0.33$ and $c_p=0.38$ mg/g.
The grey and black straight lines have a slope of $1/2.53$ and $1/3$ respectively. 
\textbf{(Inset)} Time evolution of the average size of the crystals (red dashed line) and the number of crystals 
(black line) for the same state point.
} 
 \label{fig:crystals}
\end{figure}

Now we focus on the crystallisation process. Supplementary Movie~1 shows the evolution of a 2D confocal slice. 
Even in this raw data the growth of crystalline regions from inside the liquid phase is obvious. 
More quantitatively, we detect crystalline environments by bond orientational analysis (see Supplementary Information). 
The inset of Fig.~\ref{fig:crystals} shows the evolution of the number and average size of crystallites. 
Their number first rapidly increases as nucleation events start occurring inside the liquid network, but eventually decreases as the different crystallites merge together. 
Supplementary Movie~2 shows a three-dimensional (3D) reconstruction of the whole phase ordering sequence that emphasizes crystalline particles. 
Although crystal clusters look isolated in this movie, they are bridged by the dense colloidal liquid phase and involved in the percolated gel network, as can be seen in Fig.~\ref{fig:network}b and d. 
In Fig.~\ref{fig:crystals} we show the gyration radius of individual crystalline nuclei. 
The results show that crystal growth follows two different scaling laws: at small crystalline sizes it scales with the fractal (or effective) dimension close to random percolation ($D=2.53$), while at large sizes it scales as $D=3$, as in compact crystal growth, consistently with the analysis of the structure factors, Fig.~\ref{fig:late_structure}a. 
This demonstrates that crystal nucleates inside the liquid branches, but once a nucleus reaches the transverse size of the branch, the growth proceeds in volume. Contrary to Refs~\cite{soga1999metastable,fortini2008crystallization,perez2011pathways}, thus, crystal growth is not limited by the network shape.


\begin{figure*}
 \centering
 \includegraphics{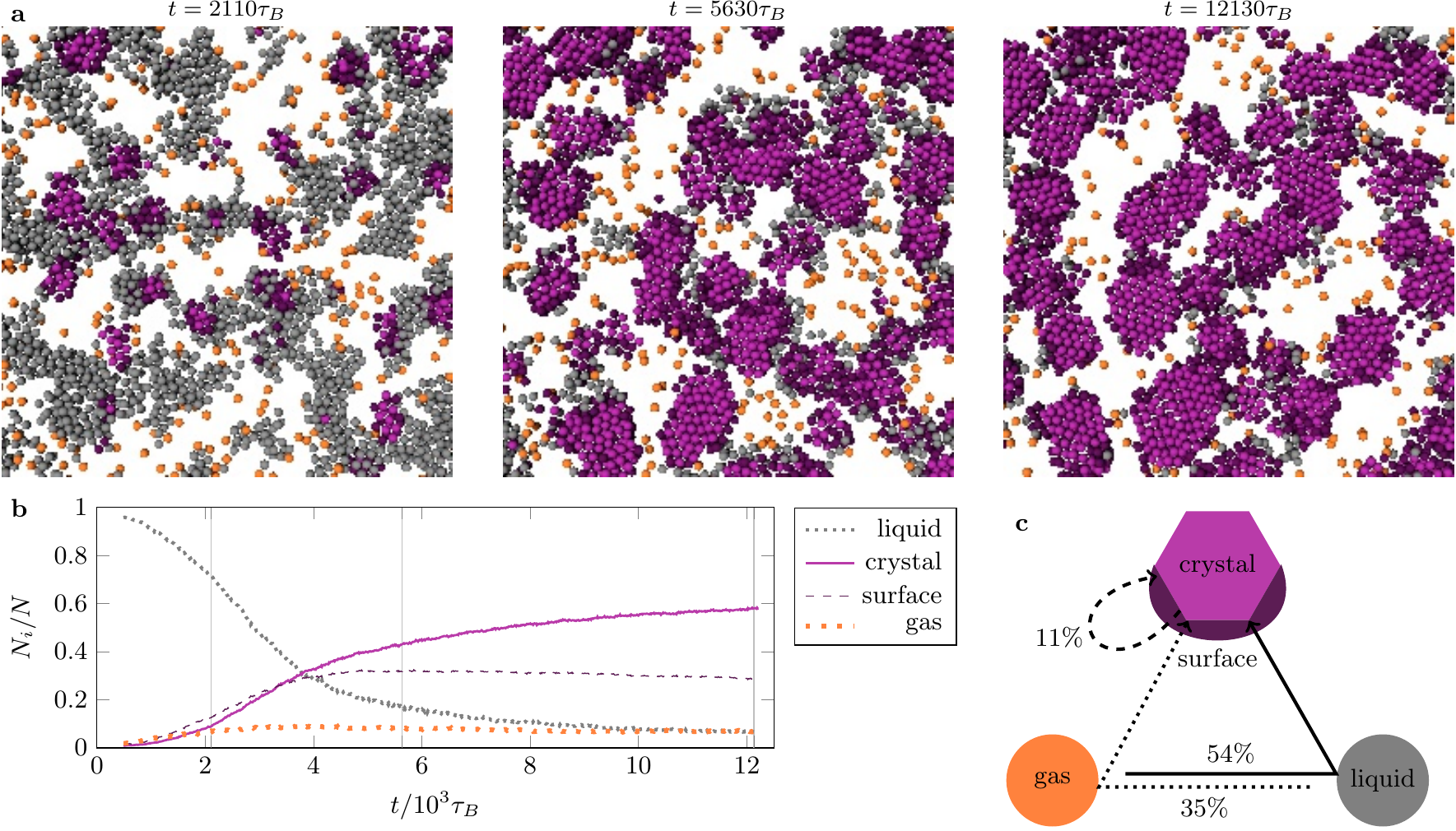}
 \caption{{\bf Crystal gel formation.}  
{\bf a,} Reconstructions from confocal coordinates at $\phi\approx 0.33$ and $c_p=0.38$ mg/g. The depth of view is 9.0 $\mu$m, while the lateral dimension is 41 $\mu$m. Particles are drawn to scale and coloured according to their phase. Purple: crystal; dark purple: crystal surface; grey: liquid; orange: gas.
{\bf b,} Fraction of particles for each phase. Gredy vertical lines indicate the times shown in panel a. 
{\bf c,} Sketch of the three routes by which a particle can end up in a crystal state. 
Probabilities are computed from the history of each single trajectory. 
A trajectory without liquid evaporation or crystal sublimation is counted as direct crystallization (continuous arrow, 54\%). 
If a liquid particle evaporates and then de-sublimates it counts in the Bergeron process (dotted arrow, 35\%). 
If sublimation is followed by de-sublimation it is Ostwald ripening (dashed arrow, 11\%). 
Most trajectories proceed via the surface state.}
\label{fig:transitions}
\end{figure*}

In Fig.~\ref{fig:transitions}a we display slabs of 5 particle diameter thickness at different times after stress driven ageing is complete. 
Particles are coloured according to their phases. 
Left panel shows the first nucleation events inside the liquid network. 
At the same time liquid regions that have not crystallized start evaporating, and the gas phase contributes to the growth of crystal nuclei (middle panel). 
Finally the different nuclei coalesce (right panel). 
Note that the topology of the crystal network found at late times, i.e. $t=5630\,\tau_{\rm B}$ (middle panel) and $t=12098\,\tau_{\rm B}$ (right panel) in Fig.~\ref{fig:transitions}a, differs significantly from that of the liquid network at early stages, i.e. $t=2055\,\tau_{\rm B}$ (left panel) in Fig.~\ref{fig:transitions}a. This suggests that the crystal network does not form only by direct freezing of the liquid, but also by other mechanisms which involve the evaporation of the liquid (Bergeron process) and the sublimation of small crystals (Ostwald ripening). In the following we will investigate these different crystallization mechanisms in depth.

In Fig.~\ref{fig:transitions}b, we show the fraction of particles in each different phase for the crystallising sample, after the gas-liquid phase separation has occurred.
The process of crystallization is characterized by a steep decrease in liquid particles, as they transform into small crystalline nuclei inside the liquid domains. 
This decrease is then accompanied by an increase in the fraction of gas particles: as the first crystals start to reach the gas phase, liquid particles evaporate to the gas phase due to the higher vapour pressure of the liquid phase compared to the crystalline phase. 
The crystalline nuclei are initially composed mostly of surface particles, but as the nuclei grow they become more compact and merge together such that the number of surface particles slowly decreases, while bulk crystals keep increasing.

After the onset of a steady-state gas population, there are three growth routes for the crystal. 
Direct crystallization is the process by which crystals grow by incorporating nearby liquid particles. 
In the Bergeron process, liquid droplets first evaporate and the resulting gas phase contributes to the growth of crystalline regions. 
Ostwald ripening is instead the process by which small crystallites sublimate, and colloidal particles are transferred to larger nuclei.
The Bergeron and Ostwald ripening processes both take place only when the crystallites are surrounded by the gas phase due to the non-existence of a liquid-crystal coexistence.
Since we have access to individual particle trajectories, we can directly assess the relative importance of these three growth channels (see Methods).
The different crystallization routes are depicted in the diagram of Fig.~\ref{fig:transitions}c.
Direct crystallization accounts for 54\% of particles trajectories in which gas or liquid particles transition to the crystal state without liquid evaporation or crystal de-sublimation. 
The Bergeron process accounts for 35\% of particle trajectories in which liquid particles transition to the gas state before crystallizing. 
Ostwald ripening accounts for 11\% of particle trajectories in which crystal particles transition to the gas state before returning to the crystal state.
In \FigProba, we confirm these results with a detailed analysis of transition probabilities.
While the direct freezing of the fluid represents the major contribution to the nuclei growth, the kinetic path via the gas phase (gas$\rightarrow$crystal), also plays a crucial role, especially in determining the morphology of the porous crystal, as we discussed above. 
In this context the Bergeron process~\cite{glickman2000glossary,morrison2012resilience} plays a considerably more important role than Ostwald ripening and is responsible for the beaded network morphology.

To summarize our results, the first stages of gelation always involve spinodal decomposition with the formation of liquid network by viscoelastic gas-liquid phase separation. 
There are then two possible arrest mechanisms:
(a) Crystallization: provided stress-driven coarsening of the network, small crystalline nuclei appear inside the liquid phase, reach the surface of the liquid branches, and then grow by addition of particles from the gas phase. 
The final structure is a network of crystal droplets, as confirmed by the fractal dimension of the branches, the volume fraction of the particles within the branches, and bond orientational analysis. 
(b) Dynamic arrest: particle arrest when the dynamics inside the liquid branch becomes slow, which should happen at the intersection of the  glass line with the liquid side of the coexistence curve.

The physical conditions required for the crystal-gel formation revealed in our study indicates that the extent of the $\phi$-$c_p$ region where mechanism (a) is operative can be widened by changing the location of the critical point, which in our system is controlled by the size of the polymer: moving the critical point to lower polymer concentrations opens the window where bonds can rearrange before the intervening glass transition. So we argue that the region of formation of crystal-gel porous structures is not only easily accessible, but also controllable.

The process of formation of ``crystal gels'' may be generic to many other 
systems. The requirements are 
(i) the presence of gas-liquid phase separation below the melting point of a crystal, 
(ii) weak or little frustration against crystallization (in our case, the use of monodisperse colloids), 
(iii) dynamical slowing down in a supercooled liquid state, which is necessary to induce viscoelastic phase separation leading to the formation of a network structure of the minority liquid phase, and 
(iv) low enough degree of supercooling to allow bond-breaking events that avoid the vitrification of the liquid phase.
Many monoatomic and single-component molecular systems can satisfy all these conditions in a certain range of the temperature and pressure. 
For example, a Lennard-Jones liquid, which represents many molecular systems without specific directional interactions, display the right kind of phase diagram~\cite{lodge1997brownian}.

Usually, monoatomic systems are very poor glass-formers and thus have not been expected to form gels.
However, our mechanism provides a novel kinetic pathway to spontaneously form network or porous structures made of crystals.
For monoatomic systems such as noble metals, condition (iii) may not be satisfied easily. 
To access glassy slow dynamics in a liquid phase and satisfy condition (iii) deep quench at high pressure may be needed. 
However, we note that even without strong dynamic asymmetry due to glassiness, bicontinuous phase separation can take place between 35-65 volume \% of the liquid phase in ordinary gas/liquid phase separation~\cite{onuki2002}; 
and, thus, porous crystalline structures can be formed, although a thin network structure may be more difficult.

Since crystals outgrow from the liquid network, well-ordered crystal planes appear on the surface of the porous structure, which is crucial for applications.
For example, nano-porous crystals of noble metals have special functions associated with ultra-high interfacial area and connectivity of pores, which are relevant to catalytic, optical, sensing, super-capacitor, and filtration~\cite{ding2004nanoporous, ding2009nanoporous, wittstock2010nanoporous,fujita2012atomic}. 
Usually such nano-porous materials are formed via at least two steps: for example, phase separation and dealloying of a mixture~\cite{erlebacher2001evolution}. 
Our novel mechanism allows us to spontaneously form sponge-like nano-porous crystals \emph{in a continuous process}, which may have an impact on many applications. 
We note that laser ablation of metals is a promising method for this purpose~\cite{povarnitsyn2013mechanisms}.
So we believe that crystal gels are an important class of heterogeneous non-ergodic states in nature and industrial applications, although they have not attracted much attention so far.

The Bergeron process is also the primary mechanism for the formation of rain drops in clouds~\cite{glickman2000glossary,morrison2012resilience}.
Our system can then be regarded as a colloidal analogue for this important process, which, for the first time, we can observe at the single-particle level.
We also speculate that our scenario could play a role in the formation of crystal networks observed in dynamically asymmetric mixtures, which includes magma~\cite{philpotts1998role}, biominerals~\cite{rousseau2005multiscale}, and foods~\cite{deman1987fat}.

\bigskip
\noindent
{\bf Acknowledgements} 
This study was partly supported by Grants-in-Aid for Scientific Research (S) (Grand No. 21224011) and Specially Promoted Research (Grand No. 25000002) from the Japan Society for the Promotion of Science (JSPS).
Collaboration between M.L. and H.Ta. has been funded by CNRS through Projet international de coopération scientifique No~7464.

\medskip
\noindent
{\bf Author Contributions} 
H.Ts. and J.R. contributed equally to this work. 
H.Ta. conceived and supervised the project, H.Ts. performed experiments, J.R. analysed the data, M.L. linked experiments and analysis, and all the authors discussed and wrote the manuscript. 

\medskip
\noindent
{\bf Additional information} 
Correspondence and requests for materials should be addressed to H.Ta. 

\medskip
\noindent
{\bf Data Availability.}
The data that support the plots within this paper and other findings of this study are available upon reasonable request.

\medskip
\noindent
{\bf Competing financial interests}
The authors declare no competing financial interests

\section*{METHODS}

\subsection*{Experiments}

We use \textsc{pmma} (poly(methyl methacrylate)) colloids sterically stabilized with methacryloxypropyl terminated \textsc{pdms}(poly(dimethyl siloxane))~\cite{klein2003} and fluorescently labelled with rhodamine isothiocyanate chemically bonded to the \textsc{pmma}~\cite{bosma2002}. To allow electrostatic repulsion, \textsc{pmma} was copolymerized with 2~\% of methacrylic acid monomers.
The diffusion constant in dilute conditions without polymer allows us to estimate the colloid diameter to $2.3\pm 0.05\, \mu$m. The Brownian time is $\tau_{\rm B} \approx 2.3\,$s. 
We assess that the size distribution of our particle is Gaussian with a polydispersity below 5\% via direct confocal measurements~\cite{Leocmach2013}.
This small polydispersity allows crystallization.
We disperse the particles in refractive index and density matching mixture of cis-decalin (Tokyo Kasei) and bromocyclohexane (Sigma-Aldrich). 
The precision of density matching is $\sim 10^{-4}$ g/ml, for which the gravity effect on our gelation processes is negligible up to 12 hours.

To induce short-ranged depletion attraction, we use polystyrene (TOSOH) as non-adsorbing polymer of molecular weight 3.8 MDa.
Experiments are conducted at 27 $^\circ$C, some 80 $^\circ$C above the theta temperature in this solvents mixture~\cite{Royall2007}. A Flory scaling of the measurements of~\cite{lu2008gelation} yields a radius of gyration $R_g=76\pm5$ nm.

Matching phase boundaries often leads to better determination of volume fractions than colloid diameter measurements~\cite{poon2012}. We obtain the best match between theoretical phase diagram (which we calculated from generalised free volume theory~\cite{Fleer2008}) and experimental data for $\sigma=2.21\,\mu$m and $R_g=80$ nm (see Supplementary Method). Therefore the polymer-colloid size ratio is $q_R=2R_g/\sigma=0.072$ and the overlap mass fraction of polymer $2.25$~mg/g. The samples displaying a gel behaviour are prepared at two different volume fractions ($\phi\approx 0.14$ and $\phi\approx 0.33$) and for different values of the polymer concentration, $c_p=0.38,0.48,0.57,1.07$ mg/g for $\phi\approx 0.33$ and $c_p=0.82,1.36$ mg/g for $\phi\approx 0.14$.

In the absence of salt, the Debye length is expected to reach a few micrometers and the (weakly) charged colloids experience a long range electrostatic repulsion. We confirm that colloids never come close enough to feel the short-ranged attraction. Screening by tetrabutylammonium bromide (Fluka) at saturated concentration brings down the Debye length to about 100~nm practically discarding the repulsion. 
Thus, salt introduction can screen the Coulomb repulsion and make the polymer-induced depletion attraction effective, initiating phase separation. 

To avoid solvent flow upon salt addition, our sample cell has two layer compartments separated by a membrane filter of pore size of 0.1 $\mu$m, permeable only to polymers and the salt ions. The top layer is 200 $\mu$m thick and contains the sample: a mixture of colloids, polymer, and solvent without salt. The bottom layer is a salt-reservoir with a half-opened structure, which allows us to exchange or insert a reservoir solution. The volume ratio of the first and second compartments is approximately 1:100.

In our experiments, the reservoir solution is initially a polymer solution without salt and electrostatic repulsion by the unscreened surface charges inhibited colloidal aggregation. Under microscopy observation, we quickly exchange the reservoir solution to a polymer solution with salt at saturated concentration (4 mM) and seal the half-opened reservoir with cover glass to avoid evaporation of solvent. Salt diffuses into the sample cell and the salt concentration becomes homogeneous within a few minutes, typically 2 minutes. This diffusive salt injection screens the surface charges of colloids, and thus initiates colloidal aggregation. This allows us to observe the process without suffering from harmful turbulent flow.

The data are collected on an upright Leica SP5 confocal microscope, using 532 nm laser excitation. The scanning volume is 98 $\times$ 98 $\times$ 53 $\mu$m$^3$, which contains $\sim 10^4$ colloid particles. Our spatial resolution is 192 nm / pixel. To be able to follow individual trajectories, we perform a 3D scan every 10~s ($\approx 4\tau_{\rm B}$) at early time and every 30 s later.

The crystal-gel state is reproduced in a sealed cell with all the components of the suspension pre-mixed. The colloid volume fraction is virtually identical at $\approx 0.33$ and polymer concentration is $0.40\,$mg/g. See \FigCap.

%

\subsection*{Structural analysis}

Calculation of the structure factor is done with the Hanning window function, to minimize boundary effects. To account for imprecision in particle localisation close to contact, we consider two particles bonded if their centre-to-centre distance is within 2.4~$\mu$m, which is the range of the potential ($\sigma+2R_g$) plus our worse estimate of localisation inaccuracy (0.15~pixels)~\cite{Leocmach2013}.

To detect percolation, two different methods have been employed. In the first one we detect percolation by looking at which frame the cluster size distribution has a more extended power-law decay. 
In the second technique, we simply measure the spatial extent of the largest cluster and consider it percolating when it is comparable to the size of the field of view of the microscope. 
Both methods lead to essentially identical percolation time for each state point.

Coarse-grained representation of the network is obtained by applying a Gaussian filter to the position of the particles (of width equal to the inter-particle distance) and plotting the surface which bounds the highest 20 \% values of the field. The local volume fraction is obtained by computing the Voronoi diagram of each configuration, which uniquely assigns a volume to each colloidal particle.

Classification into gas, liquid, crystal and crystal surface from bond orientational order~\cite{russo2013interplay} is detailed in Supplementary Information. Transition probabilities shown in \FigProba{}c and histories shown in Figure~\ref{fig:transitions} are obtained as follow: for every time frame, we assign each particle a state between gas, liquid and crystal. 
In order to minimize short-term fluctuations, the state of each colloidal particle is time averaged for $50\,\tau_{\rm B}$. 
We then measure either the time-dependent transition probability between two states, or the fraction of trajectories ending as crystal with different transition histories.




\section*{Supplementary material}

# Supplementary Information for
# "Formation of porous crystals via viscoelastic phase separation"

Hideyo Tsurusawa,[1] John Russo,[1,2] Mathieu Leocmach,[3] and Hajime Tanaka [a1]

[1]*Institute of Industrial Science, University of Tokyo,*
*4-6-1 Komaba, Meguro-ku, Tokyo 153-8505, Japan*
[2]*School of Mathematics, University of Bristol, Bristol BS8 1TW, United Kingdom*
[3]*Univ Lyon, Université Claude Bernard Lyon 1, CNRS,*
*Institut Lumière Matière, F-69622, VILLEURBANNE, France*

## DEFINITION OF STATES

The configurations obtained by tracking the position of colloids in the samples are analysed in order to determine whether a colloidal particle is in the crystal, surface, liquid or gas state. In the following we describe the classification criteria.

### Crystal

In order to detect crystalline particle we use bond orientational analysis, as described in Ref. [1]. A $(2l+1)$ dimensional complex vector $(\mathbf{q}_l)$ is defined for each particle $i$ as $q_{lm}(i) = \frac{1}{N_b(i)} \sum_{j=1}^{N_b(i)} Y_{lm}(\hat{\mathbf{r}}_{ij})$, where we set $l=12$, and $m$ is an integer that runs from $m=-l$ to $m=l$. The functions $Y_{lm}$ are the spherical harmonics and $\hat{\mathbf{r}}_{ij}$ is the normalised vector from the oxygen of molecule $i$ to the one of molecule $j$. The sum goes over the first $N_b(i)=12$ neighbours of molecule $i$. This choice accounts for the first coordination shell of close packed crystals (fcc or hcp). The scalar product between $q_{6,m}$ of two particles is defined as $\mathbf{q}_6(i) \cdot \mathbf{q}_6(j) = \sum_m q_{6,m}(i) q_{6,m}(j)$. For each pair $i$ and $j$ of neighbouring particles we define a *connection* if the scalar product $(\mathbf{q}_6(i)/|\mathbf{q}_6(i)|) \cdot (\mathbf{q}_6(j)/|\mathbf{q}_6(j)|) > 0.7$. A particle is then identified as crystalline if it has at least 7 connected neighbours.

The criteria outlined above are commonly used to identify crystalline arrangements in hard sphere systems.

### Surface, gas and liquid particles

Surface, gas and liquid particles are identified from the subset of non-crystalline particles depending on their neighbouring particles. A particle with at least two neighbouring crystal particles is classified as *surface*. A particle that is neither solid nor surface, and has at least four neighbouring particles is classified as *liquid*. A particle which is neither solid, surface nor liquid is classified as *gas*. In Fig. 6b, we plot the fraction of each state for the state point at $\phi \approx 0.30$ and $c_p = 0.38$ mg/g.

The distinction between gas and liquid phases is consistent with the analysis of volume fraction and composition of the gas and liquid phases just after phase separation (and before crystallization). To analyse the fraction of liquid and gas after phase separation we convert the position of particles into a continuous field smearing the position of each particle with a Gaussian field of variance 5 pixels, and then mapping the field to a cubic grid with 10 pixels spacing. In this way liquid and gas regions can be identified by an appropriate threshold of the field.

Figure S1 shows the probability distribution of the field, where two peaks clearly indicate the presence of gas/liquid coexistence. By choosing a threshold over the field of $f=0.06$ (vertical dashed line) we compute the fraction of particles in the solid and gas phases at $t = 490\,\tau_\mathrm{B}$, just after phase separation. The results are consistent with Fig. 6b, where the number fraction of particles in the liquid state accounts for more than 90% of the particles just after phase separation.

---

[a] e-mail: tanaka@iis.u-tokyo.ac.jp

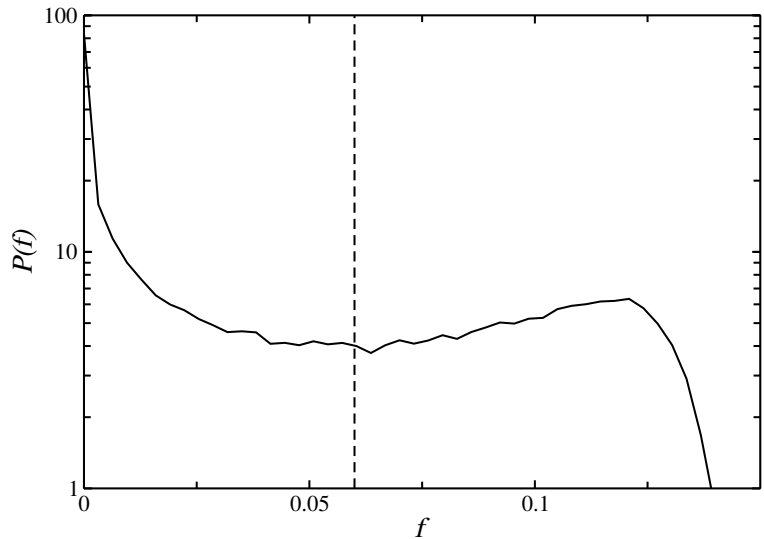

Fig. S1. **Distribution function of the Gaussian field at** $t = 490\,\tau_{\mathrm{B}}$. The Gaussian field is applied to a configuration just after phase separation and before crystallization. The vertical dashed line represents the field threshold $f = 0.06$ used to distinguish between gas and liquid particles.

## PHASE DIAGRAM

The phase diagram shown in Fig. 1is obtained in the framework of the generalized free volume theory. The colloids are considered as hard spheres and the polymers as an ideal gas. The centre of mass of a polymer cannot penetrate within $R$ of the surface of a colloid. Therefore the ideal gas is constrained to a free volume smaller than the total volume and that depends on the configuration (volume fraction) of the colloids and the polymer/colloid size ratio.

### Adjusting theoretical and experimental phase diagrams

The volume fraction of a colloidal suspension is notoriously difficult to estimate experimentally [2]. It depends on the cube of the colloid diameter and is thus extremely sensitive to any error in size determination. Therefore here we determine the volume fraction directly from the phase behaviour and deduce a precise measure of the (effective) colloid diameter from it.

Our two constraints are (i) to have the final local volume fraction of the 12-neighboured particles in the crystallising sample to be equal to the theoretical equilibrium crystal volume fraction and (ii) to have the theoretical spinodal line to correspond to the gel region boundary. Starting from our initial estimates of $\sigma$ and $R$ we iteratively converge to $R = 80$ nm, $\sigma = 2.21\,\mu$m and a crystal (without defects) at 0.723.

After this adjustment procedure, the overall agreement with our experimental points is good, except a small discrepancy at small colloidal volume fractions that is common in free volume theory [3, 4].

### Equation of state

The free volume theory relies on an equation of state (EoS) for hard spheres to deduce phase boundaries of the colloids+polymers system. Canonically, the Carnahan-Starling (CS) EoS is used for the fluid and the equation of state by Hall for the crystal. CS is accurate for a wide range of volume fraction, but lacks a divergence below 1 although a divergence of the pressure is expected at the jamming point (random close packing $\phi_J \approx 0.64$). However CS is accurate up to 0.56. Therefore CS (coupled will Hall EoS) always describes accurately fluid-crystal coexistence since the fluid volume fraction is at most 0.495.

The gas spinodal line is given by an inflexion point of the free energy, a local property of the equation of state. Therefore, the locally correct CS EoS is able to describe accurately the gas side of the spinodal region up to 0.56. The iterative process described above can be done only using CS EoS.

However, the liquid side of the binodal and spinodal lines obtained from CS EoS reach unphysically high volume fractions (above close packing), making comparisons with the measured liquid compositions impossible.

Indeed our small polymer/colloid size ratio pushes the gas-liquid critical point to high volume fractions, exemplifying the failure of CS EoS. Adding a divergence at Jamming into CS equation of state is not trivial. Attempts by Lefebre [5] and Kolafa [6] match CS at low density, have the right divergence close to jamming but far less than CS at intermediate volume fractions (0.5-0.56), the exact range where our critical point lies. Here we decided to use an equation by Liu [7] that is completely equivalent to CS up to 0.56 and then show good agreement with simulation data near jamming. The Liu EoS allows us to draw a phase diagram that is in qualitative agreement with the measured liquid compositions.

## FREE ENERGY LANDSCAPE

Following Renth et al. [8–11] we have computed, in the same framework as the phase diagram, the semi-grand potential density $\omega(\phi, \Pi)$. Here $\Pi$ is the osmotic pressure of the polymers or equivalently their chemical potential. Using the free volume theory we compute the initial value of $\Pi$ for each sample and plot the initial potential energy landscape $\omega(\phi, \Pi(t=0))$. For clarity, we subtracted a term proportional to $\phi$ irrelevant to phase behaviour so that the gas minimum and the crystal minimum appear at the same level. Results for the crystallising sample and a non-crystallising sample are shown in Fig. S2.

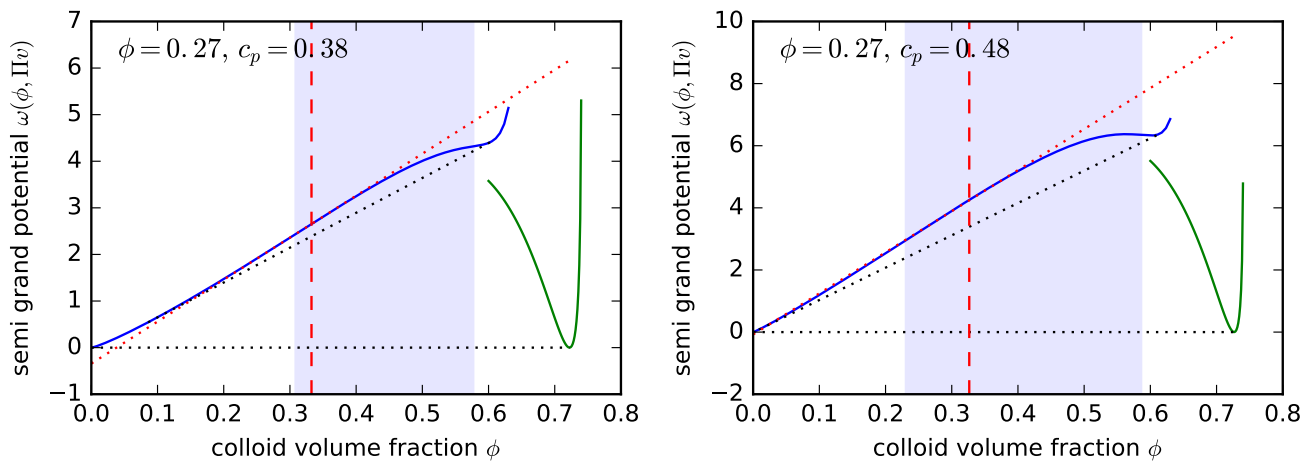

Fig. S2. **Potential energy landscape. left,** crystallising sample $\phi = 0.272$, $c_p = 0.38$ **right,** non-crystallising sample $\phi = 0.267$, $c_p = 0.48$. The two energy landscapes have the same topology. The blue and green lines are the potential for the fluid and crystal respectively. Common tangents (binodal) are drawn as dotted black lines. Note that there is no common tangent between the liquid and the crystal. The blue shaded zone is the spinodal. The red dashed vertical line shows the composition of the sample. The dotted red line is the tangent at the sample composition.

Since both samples fall in the gas-liquid spinodal region, the first step should be spinodal decomposition. However, since their is no common tangent between the liquid and the crystal, crystal nucleation should proceed from the gas in this mean-field picture. This implies the destruction of the initial bicontinuous structure and is incompatible with the growth from the liquid phase and final beaded network morphology we observe. This is contradicted by our observations.

Furthermore, there is no qualitative difference between the topology of the respective potential energy landscape of the crystallising sample and non crystallising sample. This is in stark contrast with the distinct evolutions depending on the state point we observe in our experiments. Clearly the kinetics of phase ordering in our samples cannot be predicted by free energy landscape alone, and is seriously influenced by kinetic effects coming from the momentum conservation. It is crucial to investigate the process at the particle level for accessing such information.

## TRANSITION PROBABILITIES

In Fig. S3 we plot the transition probabilities between the gas, liquid, crystal and surface states, as defined in the previous Section. Each panel represents the transition probability of a different state: **a** for the liquid state, **b** for the gas state, **c** for the surface state, and **d** for the crystal state. The results indicate that the liquid phase preferentially transforms into the surface state (as a precursor to crystallization), which accounts for the *direct crystallization channel* that was described in the main text. A high percentage of liquid transforms also in the gas phase. The gas phase itself transforms back into liquid or into surface particles, a process which is linked to the *Bergeron process*. These results support the idea that both channels are active crystallization pathways. Direct transformation of either liquid or gas into crystalline states is almost absent, indicating that our surface state indeed captures the precursor particles that attach to nuclei and later crystallize. Surface particles in fact, first transform preferentially into the liquid state, but at later time instead transform more prominently into the crystal state. The transformation of crystals into the surface state, represents the first step of the *Ostwald process*, but this channel is limited by the small rate of conversion of surface particles into the gas state.

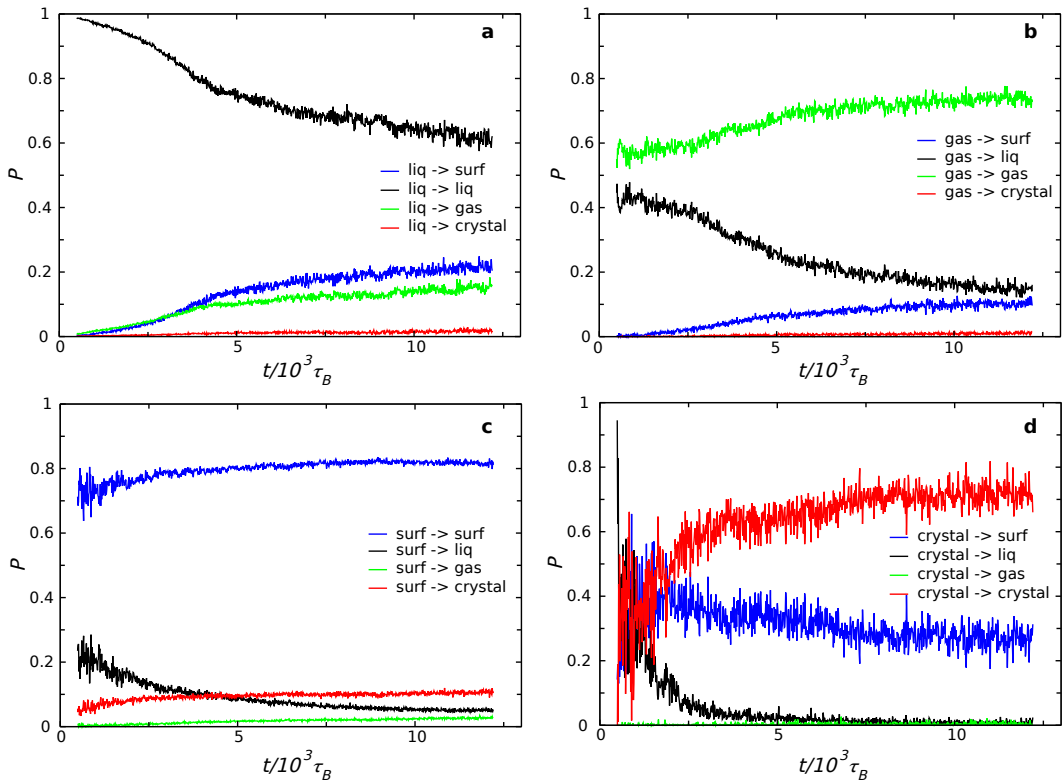

Fig. S3. **Transition probabilities during the formation of the crystal-gel. a,** Transition probabilities from the liquid state; **b,** Transition probabilities from the gas state; **c,** Transition probabilities from the surface state; **d,** Transition probabilities from the crystal state. In all panels the possible transition states are surface (blue), liquid (black), gas (green) and crystal (red). See the text for the definition of each state.

## CAPILLARY EXPERIMENT

The use of a reservoir of a salt solution through a semi-permeable membrane, allowed us to follow the entire microscopic dynamics in absence of spurious flows, from the first stages of phase separation to the formation of the crystal-gel state. In order to confirm the reproducibility of the crystal-gel state, we conducted experiments in a more traditional capillary setting, where all components of the suspension are pre-mixed in a sealed cell. A typical example is shown in Fig. S4, where a system with the colloid volume fraction of $\approx 0.33$ and the polymer concentration of $0.40\,\mathrm{mg/g}$ is considered. The figure shows the time evolution of the sample at three different times, just after phase separation ($t = 340\,\tau_{\mathrm{B}}$), and during crystal-gel formation ($t = 2600\,\tau_{\mathrm{B}}$ and $t = 6650\,\tau_{\mathrm{B}}$). The top-row shows images

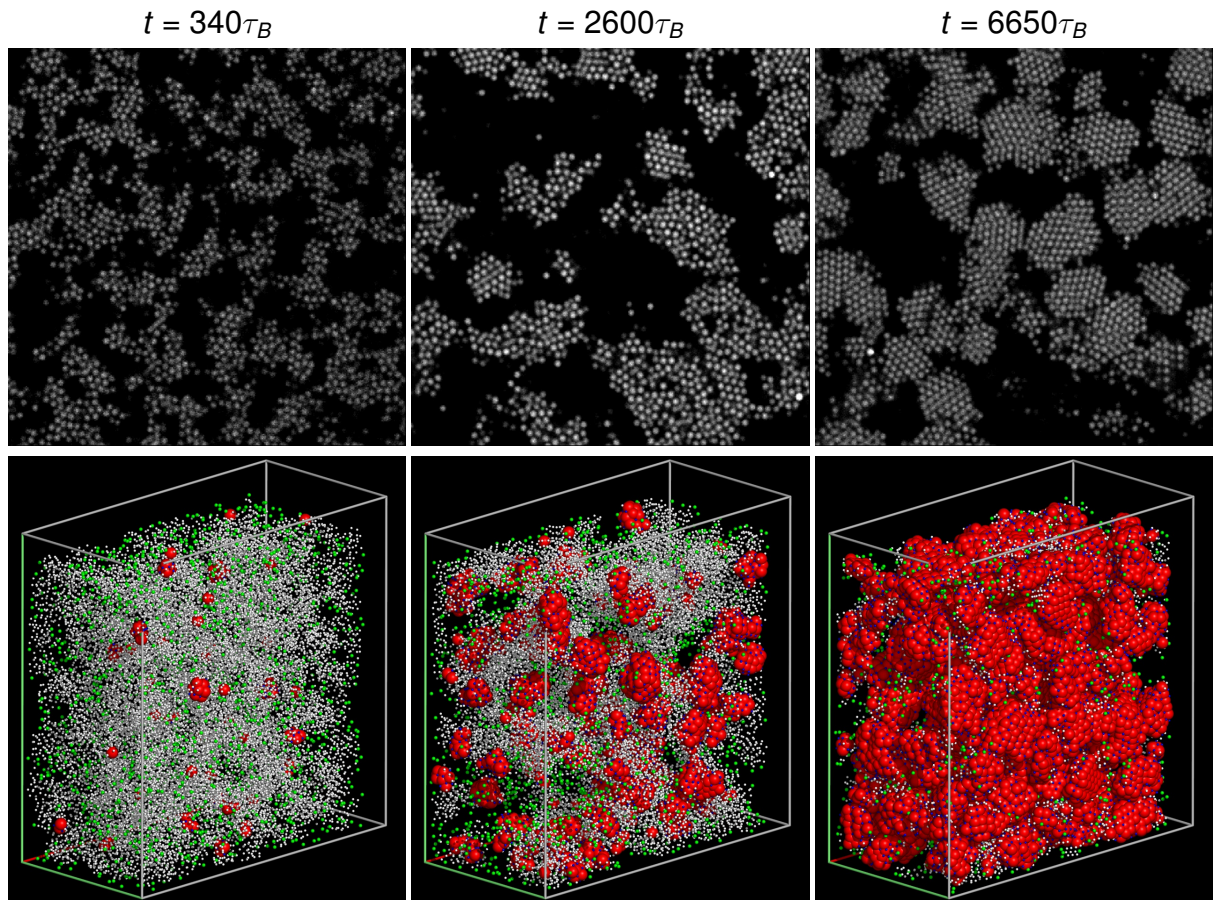

Fig. S4. **Capillary experiment reproducing the crystal-gel state.** The colloid volume fraction is ≈ 0.33 and the polymer concentration $c_p$ is 0.40 mg/g. Top row: Confocal slices showing the initial network, crystallising network and final crystal-gel network. Bottom row: Reconstructions from confocal coordinates. Particles are drawn to scale and coloured according to their phase. Red: crystal; white: liquid; green: gas; surface: blue. To highlight the crystalline regions, liquid, gas, surface particles are drawn to a tenth of the crystalline particle size.

from quasi two-dimensional confocal slices, while the bottom-row shows the three-dimensional reconstruction after single-particle tracking, where crystalline particles are highlighted in both size (large) and color (red), surface particles are depicted in blue, gas in green and liquid in white.

7

## SUPPLEMENTARY MOVIE 1

This movie shows the gelation process of $\phi \approx 0.30$ and $c_p = 0.38$ mg/g, from the original confocal images at the same z-position as Fig. 2. As stressed by the time stamps, we use two very different frame rates for the first 12 s of the movie (1160 s in real time or 530 $\tau_B$) and the remaining 15 s (27000 s in real time or 12300 $\tau_B$). We quickly started scanning the sample after exchanging the salt-reservoir with the solvent including salt. At the beginning, salt had not reached the scanning volume and colloids were dispersed by electrostatic repulsion. At time stamp $t = 0$, salt diffused into the scanning volume and screened the surface charges of colloids. Then, colloids immediately initiated phase separation by short-ranged attractions. At later times crystalline order is visible.

## SUPPLEMENTARY MOVIE 2

This movie shows the 3D structural evolution in the gelation process of $\phi \approx 0.30$ and $c_p = 0.40$ mg/g in a capillary cell. We show computer rendered trajectories obtained from the tracking of the colloidal particles. Particles are coloured according to the following code: crystal (red), surface (blue), gas (green) and liquid (white). For display purposes, crystalline particles and gas particles are respectively 400% and 50% bigger than other particles (liquid or surface).



\end{document}